\documentclass[11pt,a4paper]{article}
\pdfoutput=1
\usepackage{jheppub}
\usepackage{rotating}
\usepackage{array}
\usepackage{amsmath}
\usepackage[normalem]{ulem}
\usepackage{slashed}
\usepackage{booktabs}
\usepackage[pdftex,table,dvipsnames]{xcolor}
\usepackage{units}
\usepackage{multirow}
\usepackage{multicol}
\usepackage{url}
\usepackage{graphicx}
\usepackage{xspace}
\usepackage{subfig}
\usepackage{color}
\usepackage{diagbox}
\usepackage{bbm}

\newcommand{\epem}{\ensuremath{e^+e^-}\xspace}
\newcommand{\mupmum}{\ensuremath{\mu^+\mu^-}\xspace}

\newcommand{\ellell}{\ensuremath{l^+l^-}\xspace}
\newcommand{\hh}{\ensuremath{h^+h^-}\xspace}
\newcommand{\allpallmnoe}{\ensuremath{\mupmum/\hh}\xspace}
\newcommand{\allpallm}{\ensuremath{\ellell/\hh}\xspace}
\newcommand{\eeg}{\ensuremath{e^+e^-\to e^+e^-\gamma}\xspace}
\newcommand{\mumug}{\ensuremath{e^+e^-\to \mu^+\mu^-\gamma}\xspace}
\newcommand{\pipig}{\ensuremath{e^+e^-\to \pi^+\pi^-\gamma}\xspace}
\newcommand{\gamgam}{\ensuremath{e^+e^-\to \gamma\gamma}\xspace}
\def \belletwo {Belle\,II\xspace}

\def \superkekb {SuperKEKB\xspace}
\def \bfactory {\textit{B}-factory\xspace}

\preprint{DESY-22-023}

\title{
 \belletwo sensitivity to long--lived dark photons
} 

\author[1]{Torben Ferber,}
\author[2]{Camilo Garcia-Cely,}
\author[2]{and Kai Schmidt-Hoberg}

\affiliation[1]{Institut f\"ur Experimentelle Teilchenphysik, Karlsruher Institut f\"ur Technologie (KIT), 76131 Karlsruhe, Germany}
\affiliation[2]{Deutsches Elektronen-Synchrotron DESY, Notkestr.~85, 22607 Hamburg, Germany}

\emailAdd{torben.ferber@kit.edu}
\emailAdd{camilo.garcia.cely@desy.de}
\emailAdd{kai.schmidt-hoberg@desy.de}

\abstract{
In this letter we point out that the \belletwo experiment has a unique sensitivity to  visibly decaying long-lived dark photons. Concentrating on the signatures with a single high energy photon in association with a displaced pair of charged particles, we find that \belletwo will be able to probe large regions of parameter space that cannot be covered by any other running or proposed experimental facility. 
While the signature with charged muons or pions in the final state is expected to be background-free after all selections are applied, the case of final state electrons is more involved and requires an in-depth study.
We discuss possible ways to further suppress backgrounds and the corresponding experimental prospects.
}

\begin{document}

\maketitle

\flushbottom

\section{Introduction}

Additional $U(1)$ gauge groups naturally appear in many scenarios beyond the Standard Model and are often part of a so-called `hidden sector' \cite{Fayet:1990wx,Patt:2006fw,Batell:2009di},
which may contain the dark matter particle as well as other weakly coupled states.
Even though Standard Model (SM) particles do not necessarily possess direct charges under this U(1) factor, effective couplings between the associated $A'$ gauge boson and SM states can naturally be induced through 
kinetic mixing with the SM hypercharge field strength~\cite{Holdom:1985ag}. 
This mixing therefore provides a renormalisable portal through which $A'$ gauge bosons can be produced via the collision of SM particles, and corresponding signatures have been proposed and searched for extensively in recent years covering a very large dark photon mass range~\cite{Ahlers:2007rd,Ahlers:2008qc,Batell:2009yf,Andreas:2012mt,Essig:2013vha,Izaguirre:2013uxa,Morrissey:2014yma,Curtin:2014cca,Frandsen:2012rk,Fairbairn:2016iuf,Ilten:2016tkc,Battaglieri:2017aum,Alekhin:2015byh,Beacham:2019nyx,Batell:2014mga,BaBar:2017tiz,Banerjee:2019pds,ATLAS:2014pcp,LHCb:2019vmc,BaBar:2014zli,NA482:2015wmo}. 
In particular, light dark sectors with masses in the MeV to GeV range have triggered a lot of attention~\cite{Dolan:2017osp,Bondarenko:2019vrb,Filimonova:2019tuy,Berger:2016vxi,Biswas:2015sha,Knapen:2017xzo,Baek:2020owl,Dolan:2014ska,Krnjaic:2015mbs,Jodlowski:2019ycu} and the \belletwo experiment is known to have excellent sensitivity to many of these scenarios.

Depending on the mass ordering and coupling structure within the dark sector, the $A'$ decays invisibly into dark sector final states~\cite{Batell:2014mga,BaBar:2017tiz,Banerjee:2019pds}, visibly into SM final states~\cite{ATLAS:2014pcp,LHCb:2019vmc,BaBar:2014zli,NA482:2015wmo} or into a combination of both, leading to semi-visible final states~\cite{Izaguirre:2015zva,Izaguirre:2017bqb,Berlin:2018jbm,Duerr:2019dmv,Duerr:2020muu}.
In the following we concentrate on the case in which the $A'$ decays visibly.
For small mixing angles and small $A'$ masses the effective couplings to the SM are photon-like, so that the $A'$ decays into electrically charged SM particles.

In this letter we point out that the \belletwo experiment has a unique sensitivity to a weakly coupled visibly decaying dark photon $A'$ whose decays have a  displaced vertex. Specifically, the signature we concentrate on here is a single high energy photon in association with a displaced pair of charged particles. 
We find that the regions in parameter space that can be probed by these searches are not covered by any other running or proposed experiment and significantly extend the overall experimental reach to kinetically mixed dark photons. While
 non-electron final states will lead to signatures which can be made essentially background-free with a number of simple selections, the case of $A'\to\epem$ proves much more difficult. We discuss the problematic backgrounds in detail and 
sketch what needs to be done on the experimental side to reduce them to an acceptable level.

In the next section we briefly introduce our notation.
In section~\ref{sec:belle} we then describe the steps required to evaluate the sensitivity of \belletwo to this scenario. Our results are presented in section~\ref{sec:results} and we conclude with a short discussion in section~\ref{sec:discussion}.

\section{A brief recap of dark photons}
\label{SEC:theory}

Many theories beyond the SM exhibit additional $U(1)$ gauge factors. It is well known that the  cancellation of gauge anomalies constrain the SM charges under this $U(1)_X$ to only a few possibilities, with the simplest scenario
corresponding to vanishing charges of SM fermions. However, even in this case an effective
interaction between the mass eigenstate dark photon $A_\mu '$ and the SM particles can arise from a simple kinetic mixing in the Lagrangian, which is allowed by all symmetries. The Lagrangian including such a kinetic mixing term between the dark $U(1)_X$ and $U(1)_Y$ is given by
\begin{align}
\label{eq:LGauge}
{ \mathcal{L}} = \mathcal{L}_\text{SM} - \frac{1}{4} \hat{X}_{\mu\nu} \hat{X}^{\mu\nu} -\frac{\epsilon}{2 c_\text{W}}  \hat{X}_{\mu\nu} \hat{B}^{\mu\nu} \,,
\end{align}
together with additional terms which give rise to a mass for the dark photon - either a St\"uckelberg mass or a mass arising from a dark 
Higgs mechanism. In this letter we are agnostic about the mass generation mechanism as our results apply to both cases.

The required gauge-boson diagonalisation has been thoroughly discussed in the literature, see e.g.~\cite{Babu:1997st,Frandsen:2011cg}.  
After the fields have been diagonalised and canonically normalised one obtains the physical $Z$-boson, the SM photon $A_\mu$, and the dark photon $A^\prime_\mu$.  
For $m_{A'} \ll m_Z$ as studied in this article, the field $A'_\mu$ inherits the coupling structure of the photon to the SM fermions up to a common factor $\epsilon$. 

A comprehensive discussion of constraints and future prospects for dark photons in the mass range relevant to us has been presented in \cite{Ilten:2018crw,Bauer:2018onh,Fabbrichesi:2020wbt, Graham:2021ggy}
and we refer to those works for further details.
For our plots we adapt the limits from \cite{Bauer:2018onh} and add  limits which were subsequently published from LHCb~\cite{LHCb:2019vmc} and NA64~\cite{NA64:2019auh}, as well as projections from FASER2~\cite{FASER:2018eoc}.\footnote{We thank Patrick Foldenauer for providing the corresponding data points.}

\section{Dark photons at \belletwo}
\label{sec:belle}
In this letter we concentrate on direct production of a dark photon $A'$ in association with the SM photon $\gamma$, with the former subsequently decaying into a pair of charged SM states. This  results in a single high energy photon with a displaced pair of charged particles.
We concentrate on the case where the pair of charged particles has a significant displacement to reduce backgrounds from prompt SM processes.

Light dark photons with smaller couplings could also be targeted at a long-lived particle detector near the \belletwo interaction point 
\cite{Dreyer:2021aqd}. However, the corresponding region of parameter space is already largely constrained by beam dump experiments and hence of less interest.

\subsection{The \belletwo experiment}
%\label{sec:belle}
The \belletwo experiment at the \superkekb accelerator is a next generation \bfactory~\cite{Abe:2010gxa} that started physics data taking in 2019. 
\superkekb is a circular asymmetric $e^+e^-$ collider with a nominal collision energy of $\sqrt{s} = \unit[10.58]{GeV}$ and a design instantaneous luminosity of \hbox{$8\times10^{35}$\,cm$^{-2}$ s$^{-1}$}. The detector is described in detail in \cite{Abe:2010gxa}.

We study the \belletwo sensitivity for a dataset corresponding to an integrated luminosity of $\unit[0.5]{ab^{-1}}$, $\unit[2]{ab^{-1}}$ and $\unit[50]{ab^{-1}}$, spanning present and future prospects. 

\subsection{Event Generation}

To generate events for the process $e^+ e^- \to \gamma A'$ with subsequent decays of $A'$ we employ \texttt{MadGraph5\textunderscore{}aMC@NLO v2.7.2}~\cite{Alwall:2014hca}. 
For final states induced by the decays of $A'$ we consider $A' \to \allpallm$ ($l^{\pm}=e^{\pm}, \mu^{\pm}$ and $h^{\pm}=\pi^{\pm}, K^{\pm}$).
Given that pions and kaons behave very similarly to muons in the \belletwo detector, see e.g.\ the discussion in \cite{Duerr:2019dmv}, we do not simulate these particles explicitly. Instead, we rescale the simulated events with final state muons taking into account the relevant branching ratios, 
which we take from~\cite{Ilten:2018crw}.
As in \cite{Duerr:2019dmv} events are generated in the centre-of-mass frame with $\sqrt{s} = \unit[10.58]{GeV}$ and then boosted and rotated to the \belletwo laboratory frame.

\subsection{Signal selection}
\label{sec:selection}

\begin{figure}[tb]
 \centering
 \includegraphics[angle=0,height=8cm]{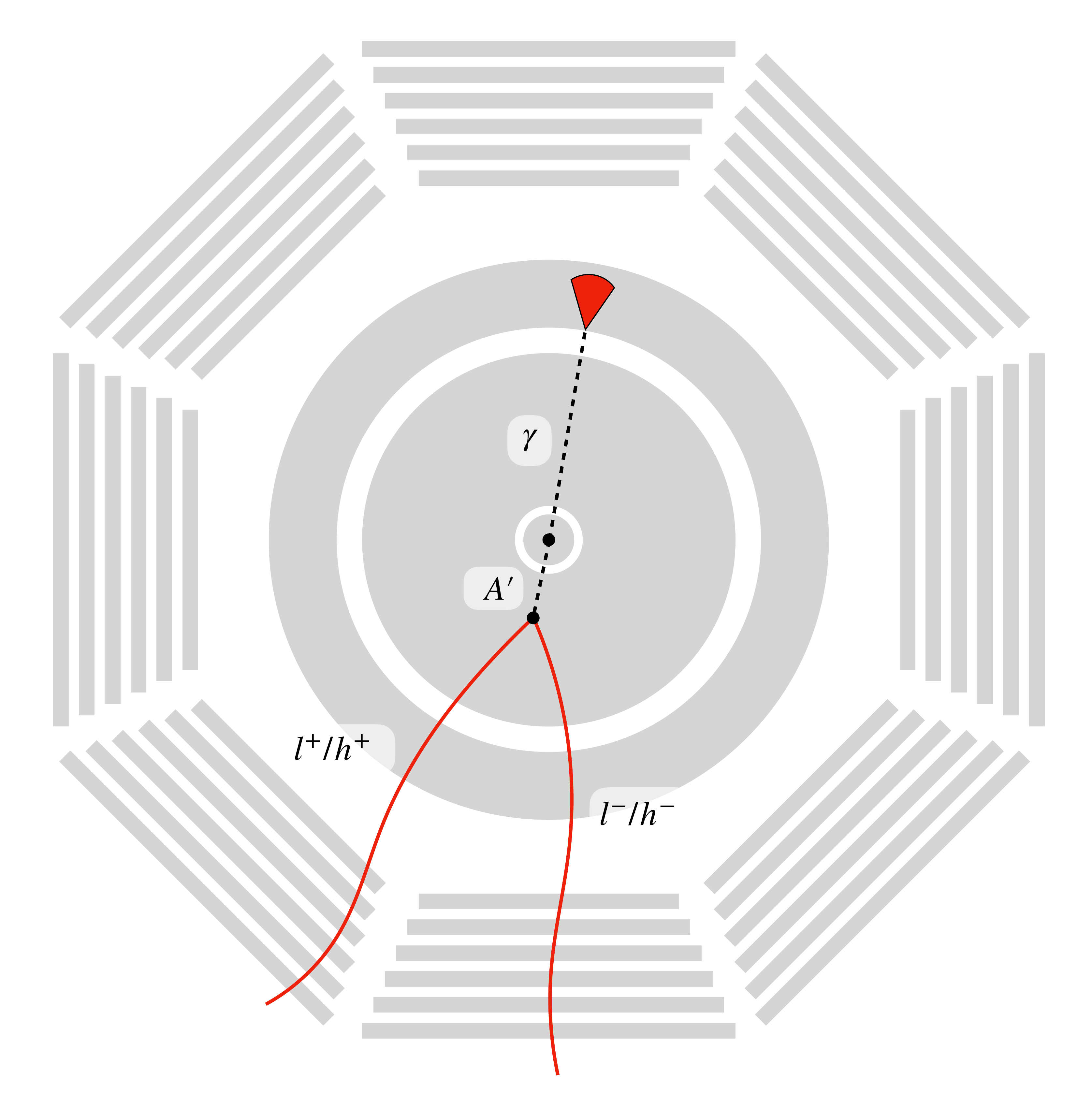}
 \caption{Schematic view of the \belletwo detector ($xy$-plane) and example displaced signature $\epem\to\gamma A', A'\to\allpallm$ ($l^{\pm}=e^{\pm}, \mu^{\pm}$ and $h^{\pm}=\pi^{\pm}, K^{\pm}$).}\label{fig:llpdarkphoton}
\end{figure}

Signal events consist of a high energy photon recoiling against a pair of high energy charged particles of opposite charge (see Fig.\,\ref{fig:llpdarkphoton}). The two charged particles originate from a common vertex and combine to an invariant mass of the dark photon $A'$. The two charged particles and the photon together combine to the collision energy of the incoming \epem beams. Experimentally, a vertex fit of the two electrons followed by a kinematic fit of all three particles will improve the invariant mass resolution significantly and allows exploiting the full potential of \epem colliders. For $A'\to\epem$ we expect that dominant background events for displaced $A'$ decays are \eeg with a misreconstructed prompt decay vertex, and \gamgam followed by a pair conversion $\gamma\to e^+e^-$; for $A'\to\allpallmnoe$ the dominant background is \mumug and \pipig with a misreconstructed prompt decay vertex, as well as the production and decay of $K_S^0$ mesons. 

As in \cite{Duerr:2019dmv} we select events based on the radial vertex position $R$ of the $A'$ decay products (\textit{region selection}), their trigger signatures (\textit{trigger selection}), and the final state kinematics (\textit{kinematic selection}). The \textit{region selections} are used to select the following vertex regions: 

\begin{itemize}
\item Decays with $R=\sqrt{x^2+y^2}<\unit[0.2]{cm}$ are very close to the nominal interaction point and will be subject to large SM backgrounds from \eeg and \mumug from vertex resolution tails and the geometrical spread of the interaction region. We assume that this region is not suitable for a displaced vertex search. However, this region is generally included in the complementary prompt searches at \epem colliders \cite{BaBar:2014zli,Kou:2018nap}.
\item The region $\unit[0.2]{cm} < R < \unit[0.9]{cm}$ is within the vacuum of the \superkekb beam pipe, but sufficiently separated from the interaction point to expect negligible backgrounds from prompt SM backgrounds \cite{Bossi:2013lxa}. While we do not expect any true conversion backgrounds $\gamma\to e^+e^-$ from \gamgam events, experience from past experiments has shown that a small fraction of conversion events that happen at larger radii will be misreconstructed with a vertex close to the nominal interaction point. Such misreconstructed tracks are often tracks that do not have hits in the inner vertex detector layers, but are extrapolated to a common vertex close to the interaction point by the tracking algorithms. We expect that requiring hits in all detector layers of the \belletwo detector downstream of the reconstructed vertex can be used to reduce these backgrounds for $A'\to\epem$ to small and maybe even negligible levels at the expense of some signal efficiency. Since the number of conversion events is very large, even small misreconstruction rates can lead to sizeable backgrounds. However, these subtle reconstruction effects must be evaluated using full detector simulations for both signal efficiency and background rejection. 
\item The region $\unit[0.9]{cm} < R < \unit[17]{cm}$ includes the beam pipe, the vertex detectors, support structures, and the inner wall of the CDC. This will induce potentially large backgrounds from $\gamma\to e^+e^-$ from \gamgam events. We expect that these backgrounds are prohibitively large for $A'\to\epem$, but negligible for $A'\to\allpallmnoe$. We therefore only consider $\mu, \pi$, and $K$ in this region. 
\item $\unit[17]{cm} < R < \unit[60]{cm}$ covers the region inside the CDC with sufficiently high tracking efficiency and only small amounts of passive material. We assume that  the backgrounds for $A'\to\epem$ are small, and that the backgrounds for $A'\to\allpallmnoe$ are negligible.
\item For $\unit[60]{cm} < R < \unit[150]{cm}$ there will be enough activity in the outer parts of the detector to veto such final states in searches for invisible final states, but not enough information to reconstruct displaced vertices. We do not include this region in our study.
\end{itemize}
In addition, an event must pass at least one of the \textit{trigger selections} that are the same as in \cite{Duerr:2020muu}. In this work we further assume that photons behave like electrons in the calorimeter-based triggers.
We note that for all $A'$ masses accessible in a displaced vertex search at \belletwo, the calorimeter cluster triggers are very efficient in triggering on the multi-GeV photon recoiling against the $A'$. 

\begin{table}[bt]
\centering
 \caption{\textit{Kinematic selections} used in our analysis. 
 \label{tab:selections}}
 \begin{tabular}{ll}
  cut on  & value \\
  \hline \hline
   \multirow{2}{8em}{decay vertex} & (i) $\unit[-55]{cm} \leq z \leq \unit[140]{cm}$\\ 
                                         & (ii) $ 17^\circ\leq \theta_\text{lab} \leq 150^\circ$ \\
   \hline
   \multirow{3}{8em}{electrons} &  (i) both $p(e^+)$ and $p(e^-) > \unit[0.1]{GeV}$\\
             &  (ii) opening angle of pair $> 0.025$ \,rad\\
             &  (iii) invariant mass of pair $m_{ee}> \unit[0.03]{GeV}$\\
             \hline
   \multirow{2}{8em}{$\mu, \pi, K$}     &  (i) both $p_\text{T}(\mu^+)$ and $p_\text{T}(\mu^-) > \unit[0.05]{GeV}$\\ 
             &  (ii)  $m_{ll} < \unit[0.480]{GeV}$ or $m_{ll} > \unit[0.520]{GeV}$ \\
   \hline
      \multirow{2}{8em}{photons}     &  (i) $E_{\text{lab}} \geq \unit[0.5]{GeV}$\\ 
                                         & (ii) $ 17^\circ\leq \theta_\text{lab} \leq 150^\circ$ \\
   \hline\hline
 \end{tabular}
\end{table}

Finally the events need to fulfil the \textit{kinematic selection} from Table~\ref{tab:selections}. The requirements on the decay vertex ensure a decay within the tracking detectors of \belletwo. For electron final states, we assume that a minimal invariant mass selection of $m_{\epem} \gtrsim 0.03\,\text{GeV}$ and a minimal opening angle of $0.025\,\text{rad}$  will be required to reject the majority of conversion backgrounds events; for the non-electron final states, we assume that particle identification alone is sufficient to reject conversions to a negligible level, but that a mass region around the $K_S^0$ mass is needed to be cut to reject backgrounds from displaced $K_S^0$ decays into two pions.

While these selections are motivated by the performance shown in \cite{Kou:2018nap}, we note that a full study of all possible backgrounds is beyond the scope of this work. 
In particular, the background for electron final states requires careful studies of reconstruction performance of the \belletwo experiment.
In a real analysis, energy resolution effects will slightly reduce the signal efficiency. We ignore this effect. We also assume that potentially increasing beam background has a negligible effect on both high momentum track resolution and high energy photon resolution~\cite{Kou:2018nap}.

\section{Results}
\label{sec:results}

\begin{figure*}[t]
\centering
\includegraphics[width=0.59\textwidth]{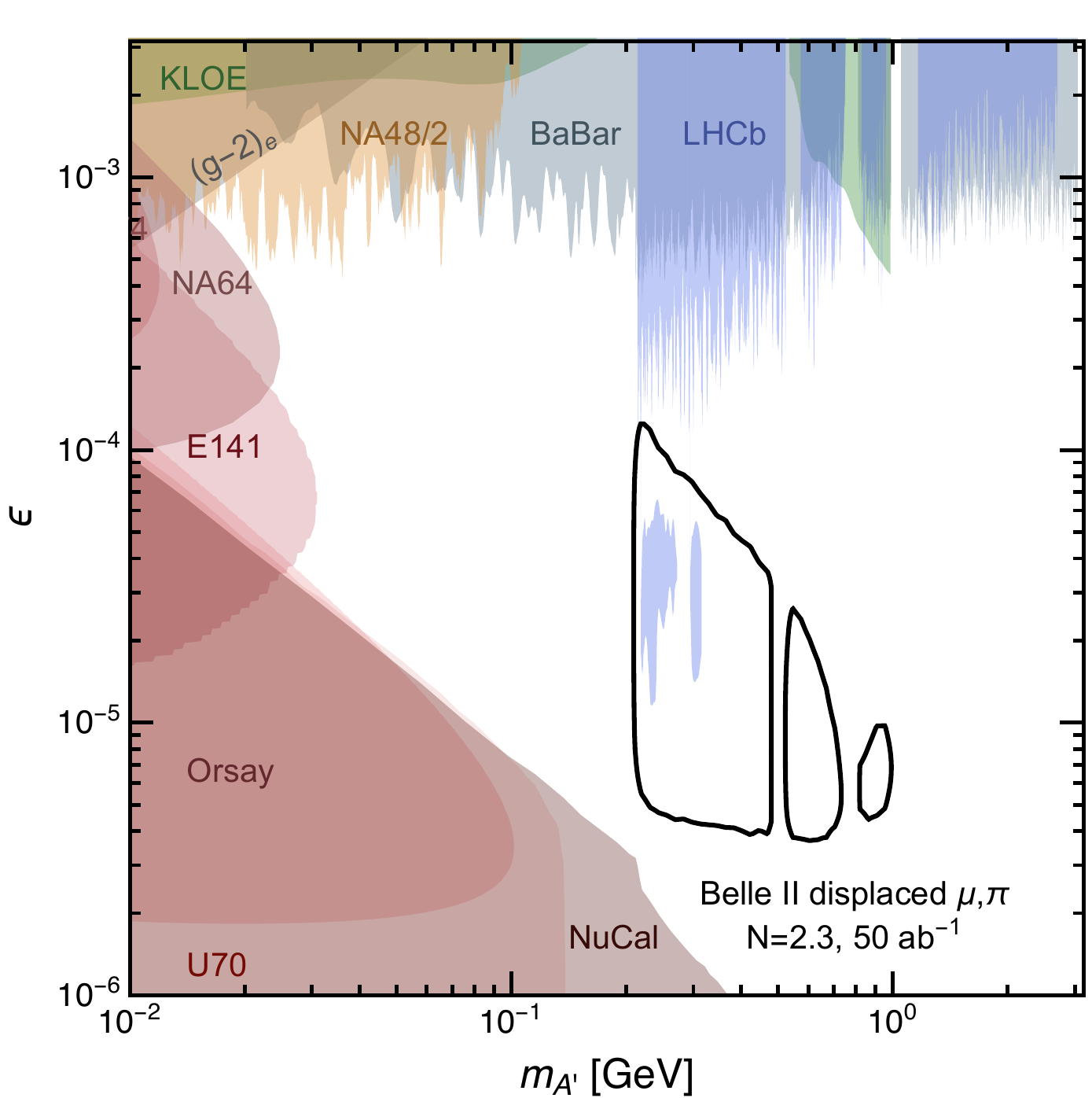}
\caption{Expected sensitivity assuming 2.3 events of the search for displaced non-electron final states at \belletwo in the $\epsilon - m_{A'}$ parameter plane for an integrated luminosity of $\unit[50]{ab^{-1}}$. Existing limits are taken from \cite{Bauer:2018onh}.  
}
\label{fig:mu}
\end{figure*}

In this section we split our results into {\it (i)} non-electron final states and {\it(ii)} final states with electrons.
For the sensitivity predictions of non-electron final states we assume zero background after selections. For the case of electrons we discussed sources of background which will not be fully eliminated by the requirements above and show a number of different sensitivities for different assumptions.

\begin{figure*}[t]
\centering
\includegraphics[width=0.49\textwidth]{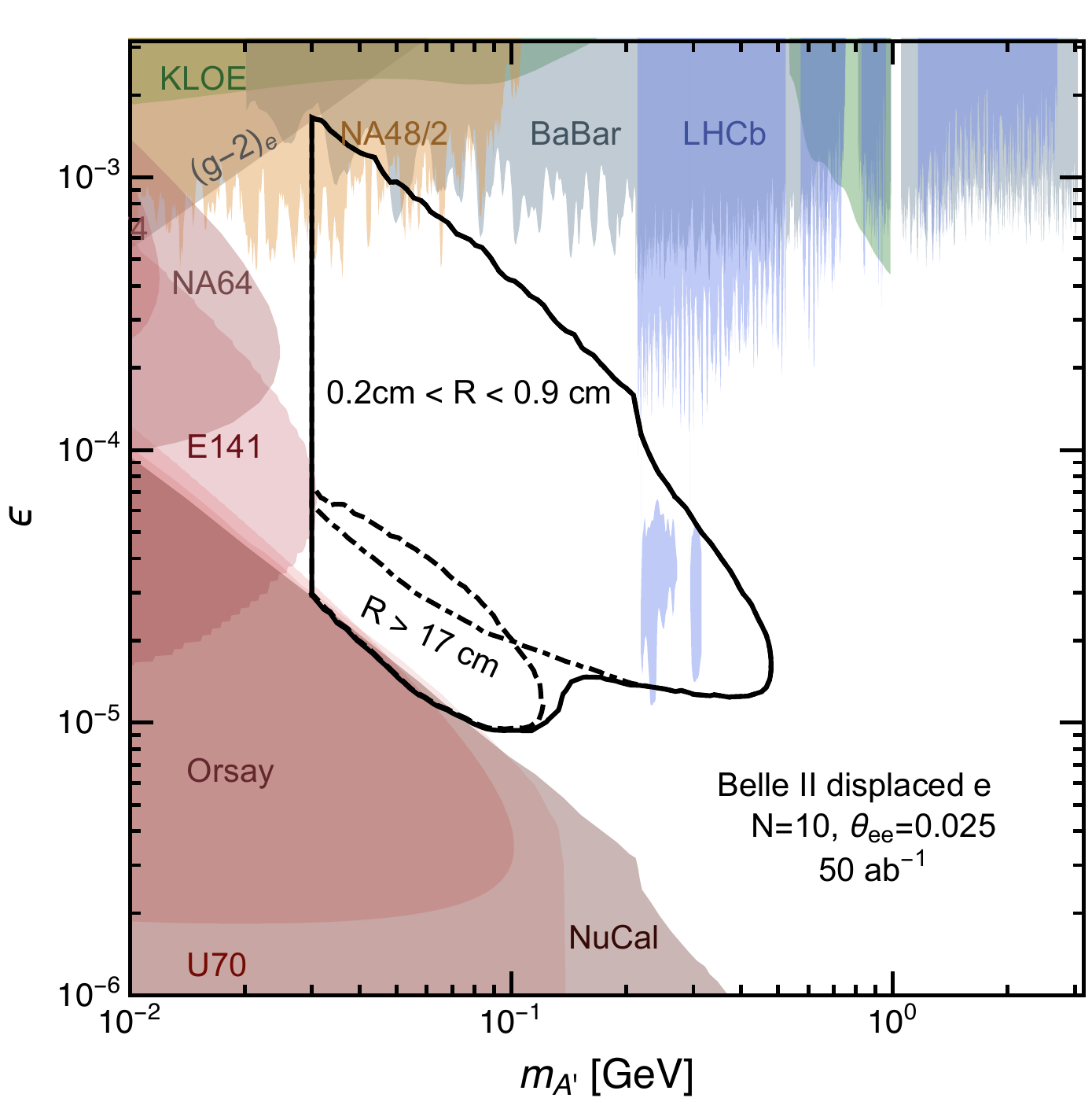}\\
\includegraphics[width=0.49\textwidth]{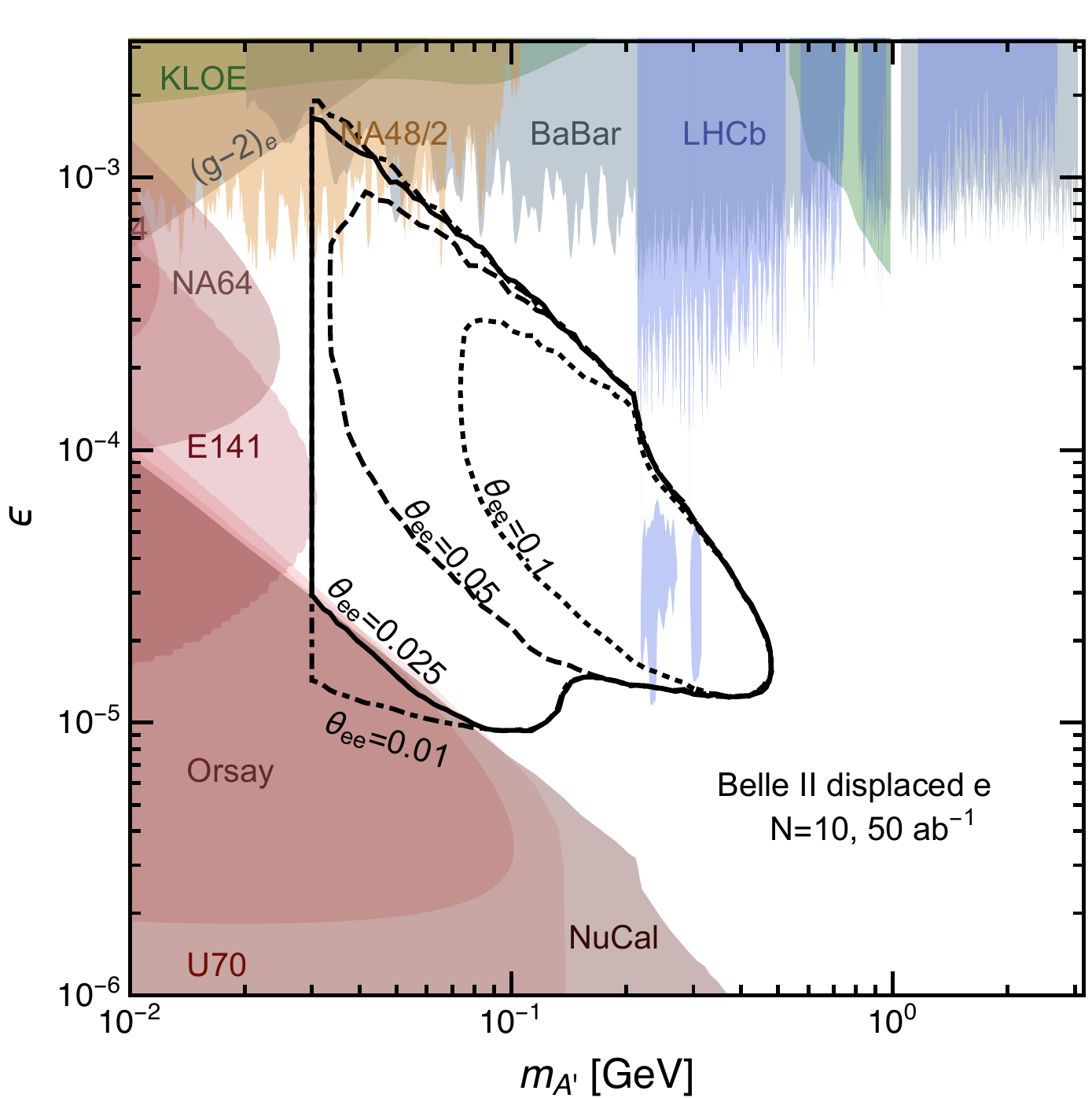}
\includegraphics[width=0.49\textwidth]{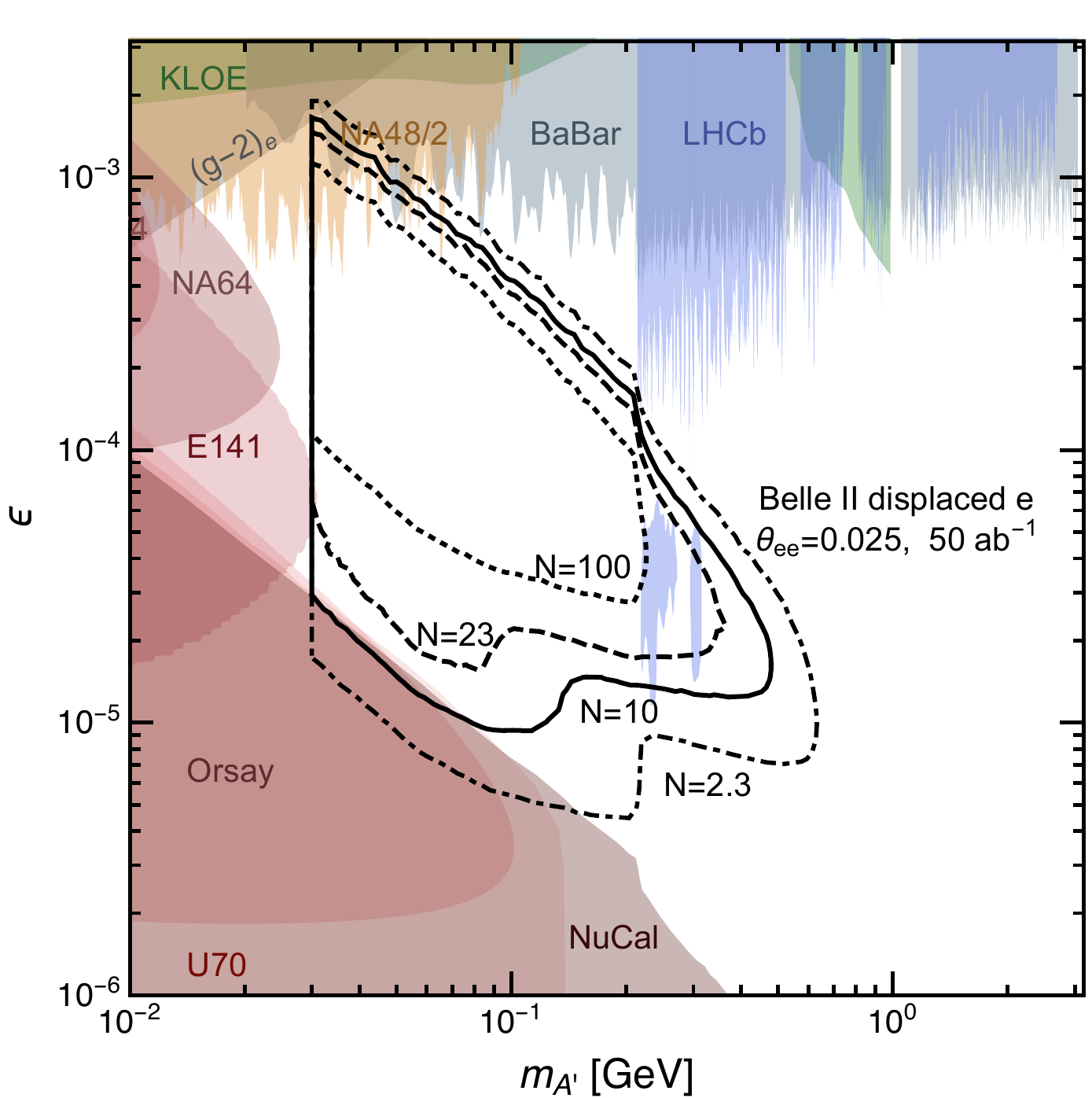}
\caption{Expected sensitivity of the search for displaced electron final states at \belletwo in the $\epsilon - m_{A'}$ parameter plane for an integrated luminosity of $\unit[50]{ab^{-1}}$. In the top panel we show the individual sensitivities corresponding to the regions in which $\unit[0.2]{cm}<R<\unit[0.9]{cm}$ (dash-dotted), $\unit[17]{cm}<R<\unit[60]{cm}$ (dashed) and the sensitivity for both regions combined (solid) for 10 events and an angular cut $\theta_{ee}=0.025$. In the bottom panel we show the sensitivity for 10 events and different angular cuts $\theta_{ee}$ (left) and a varying number of events for $\theta_{ee}=0.025$ (right). Existing limits are taken from \cite{Bauer:2018onh}.
}
\label{fig:e}
\end{figure*}

\subsection{Sensitivity for non-electron final states}
Our main result for the non-electron case is presented in Fig.~\ref{fig:mu}, where we show the expected sensitivity of \belletwo to a signature with a single high energy photon in association with displaced non-electron final states in the $\epsilon - m_{A'}$ parameter plane for an integrated luminosity of $\unit[50]{ab^{-1}}$ (black solid line) and zero backgrounds. We observe that this search is expected to cover a dark photon mass range $2 m_\mu \lesssim m_{A'} \lesssim 1$~GeV.\footnote{While we include charged kaon final states in our analysis, it is evident from this mass range that they do not contribute significantly to the sensitivity, which is dominated by muon and pion final states.} 
We find that both regions $\unit[0.2]{cm}<R<\unit[0.9]{cm}$ and $\unit[0.9]{cm}<R<\unit[17]{cm}$ contribute similarly to the sensitivity, while larger displacements do not add additional sensitivity. In total we expect at most around 50 events for $\unit[50]{ab^{-1}}$. As a result, \belletwo will become sensitive to this signature only with an integrated luminosity exceeding $\unit[2]{ab^{-1}}$.

\begin{figure*}[t]
\centering
\includegraphics[width=0.79\textwidth]{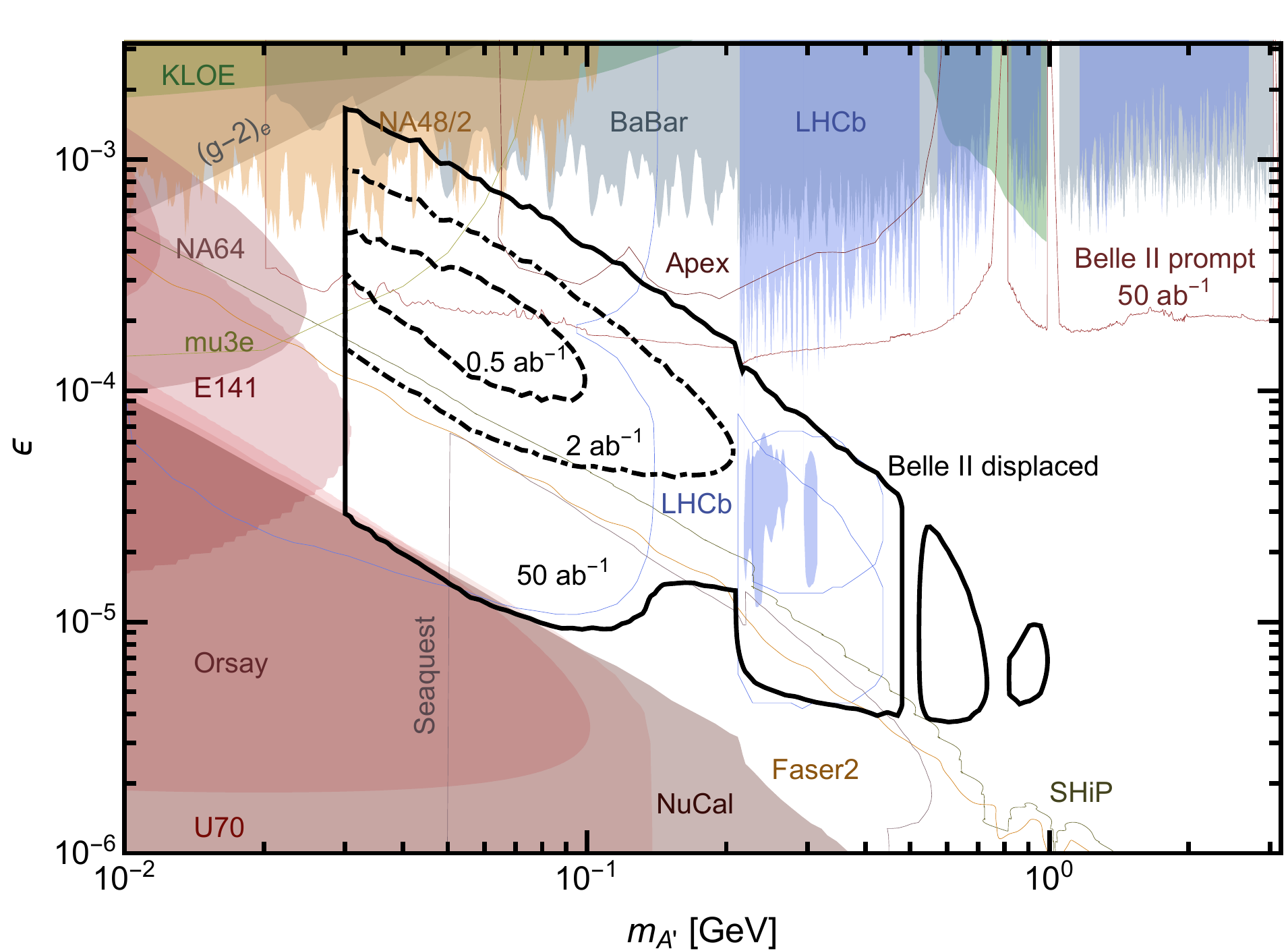}
\caption{Overall expected sensitivity of the displaced search at \belletwo in the $\epsilon - m_{A'}$ parameter plane compared with other projections assuming an integrated luminosity of $0.5\, \text{ab}^{-1}$ (dashed), $2\, \text{ab}^{-1}$ (dashed-dotted), and $50 \,\text{ab}^{-1}$ (solid). Existing limits as well as future projections are taken from \cite{Bauer:2018onh}. The \belletwo projection for the prompt $A'$ decays is taken from~\cite{Kou:2018nap}. 
}
\label{fig:combined}
\end{figure*}

\subsection{Sensitivity for electron final states}
The electron final states are experimentally significantly more challenging than the muon final states. Pair conversion events are expected at invariant masses well below 1~MeV.  
We assume that an invariant mass selection of $m_{ee}\gtrsim30$\,MeV is needed to reject the largest fraction of correctly reconstructed pair conversions due to finite tracking resolution effects. Additional selections on the opening angle $\theta_{ee}$ of the daughter electrons are likely needed to increase the tracking efficiency and momentum resolution. In the top panel of Fig.~\ref{fig:e} we show the individual sensitivities corresponding to the regions in which  $\unit[0.2]{cm}<R<\unit[0.9]{cm}$, $\unit[17]{cm} <R< \unit[60]{cm} $, and the sensitivity for both regions combined for an angular cut $\theta_{ee}=0.025$. 
We observe that a large fraction of the overall sensitivity is due to displacements below $\unit[0.9]{cm}$, but also large displacements contribute.
As outlined in Sec.~\ref{sec:selection}, we expect that additional backgrounds from misreconstructed pair conversions \gamgam cannot be reduced to zero by the selections above.
Stronger angular cuts will further reduce backgrounds and we show the sensitivity for different angular cuts $\theta_{ee}$ in Fig.~\ref{fig:e} (bottom-left).
For more optimistic and more pessimistic assumptions on the background levels, we also show the contours for 2.3, 10, 23, and 100 signal events after all selections in Fig.~\ref{fig:e} (bottom-right). This corresponds to 90\%~C.L.\ sensitivity assuming Poisson statistics for 0, and approximately 50, 300, and 6000 background events in the signal region. As discussed in Sec.~\ref{sec:selection}, additional selections may be developed to reduce background from misreconstructed pair conversions which in turn will reduce the signal efficiency. Specifically the sensitivity for 10 events could also be read as an additional reduction of signal efficiency by a factor of 0.23 to reach zero background events.

\subsection{Combined sensitivity}
In Fig.~\ref{fig:combined} we show the combined sensitivity taking into all displaced final states, assuming $\theta_{ee}=0.025$ and 10 events for the electron final state as an optimistic estimate. In addition we show all relevant projections of other future experimental facilities for comparison. We see that \belletwo not only covers a large region of parameter space which is currently still unexplored, but is also uniquely sensitive in particular for dark photon masses close to 1 GeV. As we do not expect any background events in this region, the underlying experimental uncertainties are rather small and our projection therefore correspondingly robust.

\section{Discussion}
\label{sec:discussion}
A kinetically mixed dark photon is one of the simplest and most well motivated options to extend the SM of particle physics and has attracted a huge amount of interest over the past couple of years. 
In this letter we point out that \belletwo will be able to probe a significant part of the uncovered parameter space by looking for events with a single high energy photon and a pair of displaced charged particles. The non-electron final states are experimentally rather straight forward; the electron final states require more detailed experimental studies to reject potentially large backgrounds from misreconstructed pair conversions. We find that while generally a large integrated luminosity will be required to achieve a sufficient sensitivity, \belletwo may be able to start probing so far uncovered dark photon parameter space in the near future.

\acknowledgments
We would like to thank Patrick Foldenauer for comments on the draft and for providing the limits from~\cite{Bauer:2018onh} as well as updated data. We thank Christopher Hearty for discussions during early stages of this work, and Isabel Haide for comments on the manuscript. 
This work is funded by the Deutsche Forschungsgemeinschaft (DFG) through Germany's Excellence Strategy -- EXC 2121 ``Quantum Universe'' -- 390833306 and the Helmholtz (HGF) Young Investigators Grant No.\ VH-NG-1303. C.G.C. has been supported by the Alexander von Humboldt Foundation during the first stage of this project.

%%%%%%%%%%%%%%%%
%%%%%%%%%%%%%%%%
\bibliographystyle{JHEP_improved}
\bibliography{belle2_DarkPhoton}
%%%%%%%%%%%%%%%%
%%%%%%%%%%%%%%%%

\end{document}